\documentclass[a4paper,UKenglish,cleveref, autoref, thm-restate]{lipics-v2021}
\usepackage{graphicx}
\usepackage{amsmath, amsthm, amssymb}
\usepackage{booktabs}
\usepackage{tabularx}
\usepackage{algorithm}
\usepackage{algorithmic}
\usepackage{xspace}
\usepackage{empheq}
\usepackage{tikz}
\usetikzlibrary{positioning, arrows}
\usepackage{wrapfig}
\usepackage{enumitem}
\hideLIPIcs
\nolinenumbers

\newcommand{\eg}{\emph{e.g.}\xspace}
\newcommand{\etal}{\emph{et al.}\xspace}
\newcommand{\ie}{\emph{i.e.}\xspace}

\newcommand{\tool}{\texttt{FLoPS}\xspace}

\title{\tool: Semantics, Operations, and Properties of P3109 Floating-Point Representations in Lean}
\subtitle{Rutgers Technical Report DCS-TR-762}
\titlerunning{\tool: Semantics, Operations, and Properties of P3109 FP Representations in Lean}

\author{Tung-Che Chang}{Rutgers University, USA}{tc.chang@rutgers.edu}{}{}
\author{Sehyeok Park}{Rutgers University, USA}{sp2044@cs.rutgers.edu}{}{}
\author{Jay Lim}{University of California, Riverside}{jlim@ucr.edu}{}{}
\author{Santosh Nagarakatte}{Rutgers University, USA}{santosh.nagarakatte@cs.rutgers.edu}{}{}

\authorrunning{Chang, Park, Lim, and Nagarakatte}


\DeclareMathOperator{\rd}{RD}
\DeclareMathOperator{\ro}{RO}
\DeclareMathOperator{\rn}{RN}
\DeclareMathOperator{\rne}{RNE}
\DeclareMathOperator{\ru}{RU}
\DeclareMathOperator{\rz}{RZ}
\DeclareMathOperator{\sr}{SR}

\DeclareMathOperator{\Project}{Project}

\lstdefinelanguage{Lean}{
  keywords={import, inductive, structure, where, deriving, def, match, with, namespace, lemma, theorem, class, instance, noncomputable, bool, true, false, Prop, Type, Sort},
  keywordstyle=\color{blue}\bfseries,
  ndkeywords={p3109_format, Signedness, Domain, p3109, normal_p3109, canonical_p3109, float, vnum, bounded_float},
  ndkeywordstyle=\color{teal}\bfseries,
  identifierstyle=\color{black},
  sensitive=true,
  comment=[l]{--},
  morecomment=[s]{/-}{-/},
  commentstyle=\color{gray}\ttfamily,
  stringstyle=\color{red}\ttfamily,
  morestring=[b]",
  basicstyle=\small\ttfamily,
  breaklines=true,
  showstringspaces=false,
  numbers=left,
  numberstyle=\tiny\color{gray},
  stepnumber=1,
  captionpos=b,
  literate=
    {ℕ}{{$\mathbb{N}$}}1
    {ℤ}{{$\mathbb{Z}$}}1
    {≤}{{$\le$}}1
    {≥}{{$\ge$}}1
    {→}{{$\to$}}1
    {∧}{{$\land$}}1
    {∨}{{$\lor$}}1
    {⟨}{{$\langle$}}1
    {⟩}{{$\rangle$}}1
    {≠}{{$\ne$}}1
}
\begin{document}

\maketitle

\begin{abstract}
The upcoming IEEE-P3109 standard for low-precision floating-point
arithmetic can become the foundation of future machine learning
hardware and software.  Unlike IEEE-754, P3109 introduces a parametric
framework defined by bitwidth, precision, signedness, and domain.
This flexibility results in a vast combinatorial space of formats —
some with as little as one bit of precision — alongside novel features
such as stochastic rounding and saturation arithmetic.  These
deviations create a unique verification gap that this paper intends to
address.

This paper presents \tool, \textbf{F}ormalization in \textbf{L}ean
\textbf{o}f the \textbf{P}3109 \textbf{S}tandard, which is a formal
model of P3109 in Lean.  Our work serves as a rigorous,
machine-checked specification that facilitates deep analysis of the
standard.  We demonstrate the model's utility by verifying
foundational properties and analyzing key algorithms within the P3109
context.  Specifically, we reveal that \texttt{FastTwoSum} exhibits a
novel property of computing exact ``overflow error'' under saturation
using any rounding mode, whereas previously established properties of
the \texttt{ExtractScalar} algorithm fail for formats with one bit of
precision.  This work provides a verified foundation for reasoning
about P3109 and enables formal verification of future numerical
hardware and software. Our Lean development is open source and
publicly available.
\end{abstract}

\section{Introduction}
\label{sec:intro}
The upcoming IEEE P3109 standard~\cite{p3109wg:p3109:2026} introduces
a flexible framework for low-precision, binary floating-point (FP)
formats tailored for machine learning. In contrast to
IEEE-754~\cite{ieee754:2019}, a P3109 format is defined by four
parameters: bitwidth (\ie, the total number of bits), precision (\ie,
how accurately each real value can be represented), signedness (\ie,
positive or negative), and domain (finite or extended - presence of
special values).  This parametric approach results in a large space of
potential formats with distinct encoding schemes for each
configuration. Furthermore, the set of values representable by a
given P3109 configuration (value set) deviates from IEEE-754
conventions: all formats contain a single zero, single \texttt{NaN}
(not a number), and optional infinities. These features are designed
to maximize the number of real values representable with low bitwidth
encodings.
To address the limited precision and dynamic range of low bitwidth
formats, the P3109 standard provides diverse specifications for
mapping real numbers to their floating-point counterparts. P3109
expands the rounding modes supported by IEEE-754 (\ie, round-down
($\rd$), round-up ($\ru$), round-toward-zero ($\rz$),
round-to-nearest-ties-to-even ($\rne$)) with stochastic rounding
($\sr$) and round-to-odd ($\ro$), which are utilized to enhance
convergence and stability in machine learning
training~\cite{Fitzgibbon:stochastic:arith:2025}.
Unlike the standard FP arithmetic, which typically rounds to infinity
upon overflow, P3109 operations are subject to various saturation
semantics that choose between infinity and the largest representable
value with a finite magnitude.

Formal verification of IEEE-754 floating-point arithmetic is a
well-established field~\cite{boldo:flocq:arith:2011, boldo:fp-formal-proof:2017}.
Mainstream FP verification literature assumes a "reasonably large"
precision, an assumption that could fail in the context of machine
learning formats where precision may drop to as little as one bit.
Therefore, existing formalizations of floating-point arithmetic under
IEEE-754 need to be significantly modified to handle various
properties unique to P3109.

\textbf{FLoPS}. This paper seeks to develop a formalization bespoke to
P3109 that would be appealing to verify future applications of its
formats and operations. We describe, \tool, a
\textbf{F}ormalization in \textbf{L}ean \textbf{o}f the \textbf{P}3109
\textbf{S}tandard.
We chose Lean~4 for its extensive \texttt{Mathlib} 
libraries, which include the extended reals (\texttt{EReal}) needed for 
P3109's reference semantics, and its expressive tactic language.

Our primary objective with \tool is to establish a mathematically
precise ground truth for the P3109 standard, which can serve as a
reference model for future efforts involving P3109 implementations.
As P3109-compliant hardware accelerators for machine learning are 
actively being developed, a verified specification is essential for 
ensuring that RTL implementations faithfully realize the standard's semantics. 
To avoid the complexity of raw bit-level reasoning, we establish a three-way 
equivalence between bit-level encodings, our algebraic model, and the 
closed extended real value set of each format~(Section~\ref{sec:overview}), 
which serves as a natural bridge: properties proven at the 
mathematical level transfer directly to the concrete bit-patterns that 
hardware uses, and the verified projection pipeline can serve as 
a reference for equivalence checking against hardware designs.

To show our model's practical utility, we formalize the properties of
the \texttt {FastTwoSum}~\cite{dekker:fts:numer-math:1971} algorithm
in the context of P3109~(Section~\ref{sec:fasttwosum}). Under the
round-to-nearest rounding mode, \texttt{FastTwoSum} computes the exact
rounding error of FP addition using only FP operations. Through our
formalization, we discovered that \texttt{FastTwoSum} exhibits a novel
property in the presence of saturation. When saturation occurs,
\texttt{FastTwoSum} computes exactly the ``overflow error'', the
difference between the real arithmetic result and the saturated
result, regardless of the active rounding mode. We extend our analysis
of \texttt{FastTwoSum} to develop the first formal analysis of the FP
splitting technique \texttt{ExtractScalar} \cite
{rump:accsum-1:siam:2008}. Our analysis reveals that previously
established properties of \texttt{ExtractScalar} do not extend to
P3109 formats with one bit precision (Section~\ref{sec:scalar}).
The \tool framework is open source and publicly
available~\cite{flops:github}. While developing \tool, we discovered
multiple errors in the draft P3109 standard and reported it to the
P3109 working group, which subsequently have been fixed.

\section{Background}
\label{sec:background}

\textbf{The IEEE-754 standard}. Floating-point formats as defined by
the IEEE-754 standard for binary floating-point
arithmetic~\cite{ieee754:2019} represent subsets of real numbers,
augmented by special values such as $\pm \infty$ and \texttt{NaN}. The
set of finite real values in a format $f$ is determined by three
parameters: precision ($P$), the maximum exponent ($\texttt{emax}$),
and the minimum exponent ($\texttt{emin}$). Given a binary indicator
$s$ for the sign, a finite real $x$ in the value set of $f$ can be
expressed as $x = (-1)^{s} \times m \times 2^{e-P+1}$ where $m$ and
$e$ each represent $x$'s significand and the canonical exponent. For a
non-zero $x$, $e = \text{max}(\lfloor \text{log}_{2}(|x|) \rfloor,
\texttt{emin})$. If $x = 0$, $e = \texttt{emin}$. For $x$ to be a
member of $f$'s value set, its significand and exponent must satisfy
$m, e \in \mathbb{Z}$, $0 \le m < 2^{P}$, and $\texttt{emin} \le e \le
\texttt{emax}$. IEEE-754 additionally requires that if $e >
\texttt{emin}$, then $m \ge 2^{P-1}$. Based on this requirement, the
standard further divides finite reals into two categories. A non-zero
number $x$ is \texttt{normal} if $|x| \ge 2^{\texttt{emin}}$ (\ie, $m
\ge 2^{P-1}$) and \texttt{subnormal} otherwise.

\begin{wrapfigure}{r}{.3\textwidth}
    \includegraphics[width=\linewidth]{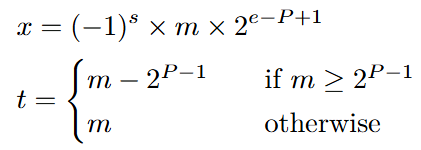}
    \includegraphics[width=\linewidth]{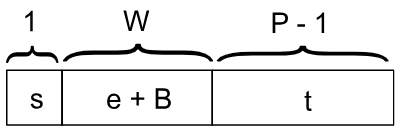}
  \caption{
    \small The binary encoding of finite floating-point numbers.
    For a finite $x$, $t$ represents the $P-1$ trailing bits of $x$'s 
    significand $m$. $B$ and $W$ each represent the exponent bias and 
    bitwidth.
  }
  \label{fig:fp-encoding}
\end{wrapfigure}

For each value in a given format, the standard assigns a corresponding 
binary encoding. An encoding contains three fields, each designated for the 
sign, exponent, and significand respectively. For a finite $x$, the significand 
field stores the $P-1$ trailing bits of $x$'s significand. Given 
$K$-bits for the full encoding (\eg, $64$ for double-precision), with $1$-bit 
and $P-1$-bits reserved for the sign and significand respectively, the exponent 
field occupies the remaining $W = K - P$ bits. Using the $W$-bits available, 
the exponent field stores the sum of $x$'s exponent and the format's exponent 
bias $B$, which is defined as $B = \texttt{emax}$ (see Figure ~\ref{fig:fp-encoding}).

Of the $2^{W}$ possible values for the exponent field, $2^{W} - 2$ are
designated for the biased exponents of normal values. The extremal
encoding $E = 0$ is reserved for 0 and subnormal numbers while
$E=2^{W}-1$ (\ie, exponent field containing only ones) is reserved for
special values.
Consequently, the maximum exponent $\texttt{emax}$ maps to the second
largest exponent encoding $E = 2^{W} - 2$. If $E = 2^{W} - 1$ and the
significand field consists entirely of zeroes, the encoding represents
an infinity, with the sign being determined by the encoding's leading
bit. Conversely, if $E = 2^{W} - 1$ and the significand field is
non-zero, the encoding represents a \texttt{NaN} value.  The value in
the significand field subsequently determines the type of the
\texttt{NaN}. Given that encodings store sign bits explicitly, another
notable property of IEEE-754 formats is that there exist two zeroes
$\pm 0$.  For any given IEEE-754 format with $K$-bit-encodings, $+0$
and $-0$ each correspond to the encodings $0$ and $2^{K-1}$.

A floating-point format $f$ can only represent numbers using $P$-bits of 
precision. However, the real arithmetic result of an operation on 
floating-point numbers may not be exactly representable with $P$ precision bits 
(\eg, $1 + 2^{-P}$). Therefore, floating-point operations must augment their 
real arithmetic counterparts with a rounding function $\circ: \mathbb{R} 
\rightarrow \mathcal{V}_{f}$ that maps real numbers to the target format $f$'s 
value set. Rounding in turn can incur errors due to loss of precision. 
To mitigate errors, a rounding function must 
ideally possess certain properties, the most notable of which are monotonicity 
and \textit{faithfulness}. Given a target value set $\mathcal{V}_{f}$ and a 
number $r \in \mathbb{R}$, the largest $v \in \mathcal{V}_{f}$ such that 
$v < r$ is $r$'s \textit{floating-point predecessor} ($pred(r)$). Similarly, 
the smallest $v \in \mathcal{V}_{f}$ such that $v > r$ is $r$'s 
\textit{floating-point successor} ($succ(r)$). Collectively, 
$pred(r)$ and $succ(r)$ are referred to as $r$'s floating-point neighbors.
A rounding function is \textit{faithful} if for any $r \in \mathbb{R}$, 
$\circ(r) = r$ when $r \in V_{f}$ and $\circ(r) \in \{pred(r), succ(r)\}$ 
otherwise.

For binary formats, the IEEE-754 standard supports four rounding modes
$\rd$, $\ru$, $\rz$, and $\rne$. All four rounding modes are faithful,
with each mode enforcing specific rules for choosing between
floating-point neighbors when a number $r$ is not representable in the
target format.  For such cases, $\rd$ always chooses $pred(r)$ while
$\ru$ chooses $succ(r)$.  A rounding function $\circ$ is thus also
faithful if $\forall r \in \mathbb{R}, \circ(r) \in \{ \rd(r), \ru(r)
\}$.  $\rz$ applies $\rd$ when $r > 0$ and $\ru$ otherwise.  $\rne$,
on the other hand, applies the round-to-nearest rule and thereby
always chooses the floating-point neighbor closest to $r$. If $r$ is
at an exact midpoint between its neighbors (\ie, $|pred(r) - r| =
|succ(r) - r|$ ), $\rne$ breaks the tie based on the floating-point
neighbors' parity. In such cases, $\rne$ as specified by IEEE-754
chooses the neighbor that has an even significand.

Given the maximum representable exponent $\texttt{emax}$, the maximum
representable finite magnitude in a IEEE-754 format is $M_{hi} = (2^{P} - 1) \times 2^{\texttt{emax} - P + 1}$. Numbers $r \in \mathbb{R}$ such
that $|r| > M_{hi}$ are subsequently in the \textit{overflow}
range. When handling overflow, the IEEE-754 standard applies rounding
using the aforementioned rounding modes to decide between $M_{hi}$ (or
$-M_{hi}$) and $\infty$ (or $-\infty$). For example, any $r$ such that
$r < -M_{hi}$ will round to $-\infty$ under $\rd$ while numbers $r >
M_{hi}$ round to $M_{hi}$.

\textbf{The upcoming IEEE P3109 standard}. The P3109
standard~\cite{p3109wg:p3109:2026} allows for the derivation of a
format $\textit{f}$ via a tuple of four parameters: bitwidth ($K$),
precision ($P$), signedness ($\Sigma$), and domain ($\Delta$). The
bitwidth $K$ specifies the total storage size and must be at least 3
bits. The precision $P$ must be at least 1. $P$ is further constrained
by the signedness parameter $\Sigma \in \{ \texttt{Signed},
\texttt{Unsigned} \}$: $P$ must be strictly less than $K$ for
$\Sigma=\texttt{Signed}$ and must be no greater than $K$ otherwise.

While IEEE-754 defines the exponent bias $B$ using the maximum
representable exponent $\texttt{emax}$, P3109 instead directly defines
$B$ using the available bitwidth and precision. For $\texttt{Signed}$
formats, $B = 2^{K-P-1}$; for $\texttt{Unsigned}$ formats, $B =
2^{K-P}$. Given the exponent field width $W$, which is $K-P$ for
\texttt{Signed} formats and $K-P+1$ otherwise, the bias of a P3109
format subsequently determines the maximum exponent. Unlike IEEE-754,
P3109 does not reserve the highest exponent field entirely for special
values such as \texttt{NaNs} and $\pm \infty$.

$\Sigma$ determines the inclusion of
negative reals in a format's value set. Likewise, the domain parameter $\Delta
\in \{\texttt{Finite}, \texttt{Extended}\}$ determines the inclusion of
infinities. Formats in the \texttt{Finite} domain represent only finite numbers
and \texttt{NaN}, while those in the \texttt{Extended} domain include
infinities. Specifically, an \texttt{Extended} format includes both $\pm
\infty$ if it is also \texttt{Signed} and only positive $\infty$ otherwise.
The bit integer encodings for these special values are parametric and vary 
across formats depending on $K$, $\Sigma$, and $\Delta$. We present the 
encodings of special values across different formats in Table~\ref{tab:special_values}.

\begin{table}[h]
    \centering
    \small 
    \caption{Encodings of special values for bitwidth $K$. N/A denotes
      that the value is not representable in the given formats. }
    \label{tab:special_values}
    \begin{tabular}{lcccc} 
        \toprule
        \textbf{Value} & \textbf{Signed Extended} & \textbf{Signed Finite} & \textbf{Unsigned Extended} & \textbf{Unsigned Finite} \\
        \midrule
        \texttt{NaN} & $2^{K-1}$ & $2^{K-1}$ & $2^{K}-1$ & $2^{K}-1$ \\
        $+\infty$ & $2^{K-1}-1$ & N/A & $2^{K}-2$ & N/A \\
        $-\infty$ & $2^{K}-1$ & N/A & N/A & N/A \\
        \bottomrule
    \end{tabular}
\end{table}

As shown in Table~\ref{tab:special_values}, each P3109 format supports
a \textit{singular} instance of \texttt{NaN}. Moreover,
\texttt{Signed} P3109 formats do not support signed zeroes. The
encoding that the IEEE-754 standard reserves for $-0$ is instead used
for \texttt{NaN}. Lastly, the infinities share the maximum exponent
field $2^{W} - 1$ with the maximum exponent finite values. Given that
the available bitwidth can be as low as 3 bits, these design choices
collectively aim to maximize the set of finite values representable
using low bitwidth encodings.

For arithmetic operations, P3109 supports a superset of IEEE-754's
standard rounding modes. P3109 additionally provides round-to-odd
($\ro$) and stochastic rounding ($\sr$), which can improve the
stability of training for machine
learning~\cite{Fitzgibbon:stochastic:arith:2025}.  Unlike
deterministic modes, $\sr$ rounds a value $x$ to its floating-point
predecessor or successor with probabilities proportional to $x$'s
distance from each neighbor, making it unbiased (\ie,
$\mathbb{E}[\circ(x)] = x$).  P3109 specifies three $\sr$ variants
that differ in how finite random bits are used: the fastest variant
that can introduce systematic bias, a fast variant that is unbiased
for infinite-precision inputs, and a corrected variant that remains
unbiased even for finite-precision inputs. While sharing several
rounding modes with IEEE-754 counterparts, P3109 formats can exhibit
unique rounding behaviors due to their low precision. For example, the
conventional definition of $\rne$ for IEEE-754 formats do not apply to
P3109 formats where $P = 1$ as their values can never have
\textit{even significands}.  For $P = 1$, parity-based rounding (\eg,
$\rne$, $\ro$) under P3109 instead relies on the parity of
\textit{exponents}.  Moreover, P3109 extends IEEE-754's
precision-based rounding with three additional \textit{saturation
  modes}, which explicitly define the rounding decision for numbers in
the overflow range.  The \texttt{SatFinite} mode clamps values in the
overflow range to the maximum representable finite magnitudes while
\texttt{SatInf} maintains the standard overflow
behavior. \texttt{SatPropagate} preserves infinities, but otherwise
enforces the same behavior as \texttt{SatFinite} for finite reals.

\section{The \tool Framework}
\label{sec:overview}
\begin{wrapfigure}{r}{.49\textwidth}
    \centering
    \resizebox{0.45\columnwidth}{!}{%
    \begin{tikzpicture}[
        node distance=4cm,
        every node/.style={font=\large, align=center},
        edge label/.style={font=\normalsize, auto, swap},
        double arrow/.style={double, <->, >=stealth, thick}
    ]
        \node (adt) at (0, 4) {\textbf{P3109 ADT}\\ $\texttt{p3109}_f$ };
        \node (bits) at (-4.5, 0) {\textbf{Bit-Level}\\\text{Fin}$(2^K)$};
        \node (sem) at (4.5, 0) {\textbf{Semantic Domain}\\$\mathcal{V}_f \subset \mathbb{R} \cup \{\pm\infty, \text{NaN}\}$};

        \draw[double arrow] (adt) -- (bits)
            node[midway, left=0.3cm, align=right, font=\normalsize]
            {\textbf{Bijection}\\$\Phi/\Phi^{-1}$};

        \draw[double arrow] (adt) -- (sem)
            node[midway, right=0.1cm, align=right, font=\normalsize]
            {\textbf{Fidelity (Round-Trip)}\\$\texttt{encode}([\![a]\!]) = a$};

        \draw[double arrow, dashed] (bits) -- (sem)
            node[midway,  align=center, font=\normalsize]
            {\textbf{Corollary: Isomorphism}\\Composition of Bijection \& Fidelity};
    \end{tikzpicture}%
    }
    \caption{\small The "Triangle of Correctness" illustrating the
      formal isomorphism between the bit-level representation, the
      algebraic model, and the semantic value set.}
    \label{fig:correctness-triangle}
\end{wrapfigure}

We describe our \tool framework for the formalization of low-precision
representations in the P3109 standard. To bridge the gap between
low-level bit-patterns used in the P3109 standard with high-level
mathematical reasoning, \tool has three primary parts: (1) bit-level
encodings, (2) an algebraic model with inductive types, and (3) a
semantic domain of value sets (\ie, closed extended reals). To
facilitate rigorous proofs, we define the P3109 formats and values as
inductive types.  This layer abstracts away from bit-level details to
enforce the standard's parametric constraints on bitwidth, precision,
and special value encoding directly in the type system.  By treating
floating-point numbers as algebraic structures rather than as
bit-patterns, we simplify the formalization of operations and
predicates.

\textbf{Three-way isomorphism}. To ensure our model is a faithful
representation of the P3109 standard, we establish a three-way formal
equivalence relation, as shown in
Figure~\ref{fig:correctness-triangle}. Specifically, it involves
showing (1) \emph{encoding equivalence} and (2) \emph{semantic
equivalence}. The encoding equivalence step ensures a verified
bijection between the bit-level encodings and our algebraic model. The
semantic equivalence step also provides a proof of bijection between
the algebraic model and the set of closed extended real values in the
value set of a given P3109 format. These two steps guarantee that any
property proven within our algebraic model automatically holds for the
actual bit-level implementation described by the standard.

\tool also includes a formalization of the core projection pipeline
that maps closed extended real values back into the P3109 domain.
This includes a bespoke implementation of rounding modes, specifically
handling the unique ``ties-to-even'' logic required for $P=1$ formats,
and the saturation semantics, which is useful to verify high-level
properties of foundational FP algorithms in
Section~\ref{sec:fasttwosum}.

\subsection{The P3109 Algebraic Model}
\label{sec:formalization}
The P3109 algebraic type is indexed by a P3109 format as detailed in 
Section~\ref{sec:background}. The core constructor of the ADT is a pair of 
integers $m$ (significand) and $e$ (exponent), supplemented with the 
constraints they should satisfy according to the index format. Note that 
because $e$ is not explicitly scaled with respect to the precision, 
$e$ represents the exponent of the significand's \textit{least significant bit}.

\textbf{Format parameters and derived values}. We define a format $f$
using a structure containing the four fields $K$, $P$, $s$ (\ie,
$\Sigma$) and $d$ (\ie, $\Delta$) along with their constraints.

\begin{lstlisting}[language=Lean, caption={Definition of the \texttt{p3109}
  format structure}]
structure p3109_format where
  K : ℕ
  P : ℕ
  s : Signedness
  d : Domain
  h_K : 2 < K
  h_P : 0 < P ∧ (s = Signedness.signed → P < K) ∧ (s = Signedness.unsigned → P ≤ K)
\end{lstlisting}

To define valid values within a format's dynamic range, we derive minimum
and maximum representable exponents from the format's bias $B$. In defining
the minimum and maximum exponents, we distinguish between exponents assigned to
the most significant bit (\ie, $m \times 2^{e_{msb}-P+1}$) and those assigned 
to the least significant bit (\ie, $m \times 2^{e_{lsb}}$).  
The \textbf{minimum exponent} (\texttt{emin}) 
for the MSB is uniformly defined as $\texttt {emin} = 1 - B $. The minimum 
exponent for the LSB is defined as $\texttt{emin}_{lsb} = \texttt{emin} - P + 1$.

\begin{wrapfigure}{r}{.5\textwidth}
  \includegraphics[width=\linewidth]{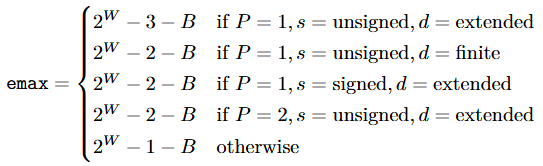}
  \caption{\small The definition of \texttt{emax} for all $K$, $P$, $s$, and $d$.}
  \label{fig:emax}
\end{wrapfigure}

Unlike \texttt{emin}, the definition of \textbf{maximum exponent} varies 
across formats due to differences in the encodings of special 
values (see Table~\ref{tab:special_values}). The maximum possible exponent for 
the MSB (\texttt{emax}) is defined via case analysis on the format parameters
and their associated exponent field length $W$.
The default definition $\texttt{emax} = 2^{W} - 1 - B$, for which the encoding
is $2^{W} - 1$, covers the majority of the formats. However, lower precision 
formats deviate from this definition. 
When $P=1$, the trailing bits of the 
binary encoding form the exponent field (\ie, no significand field). Moreover, 
the largest binary encodings are reserved for the special values. 
For example, for an \texttt{Unsigned Extended} format such that $P=1$ 
implies $W = K$, \texttt{NaN} and $\infty$ occupy $2^{W} - 1$ and $2^{W} - 2$ 
respectively. The maximum possible biased exponent in this case can be 
at most $2^{W}-3$, and thus $\texttt{emax} = 2^W-3-B$. A similar case also 
arises for the \texttt{Unsigned Extended} formats when $P = 2$. 
Figure~\ref{fig:emax} provides the full definition of \texttt{emax} covering 
all cases. Given the definition for \texttt{emax}, we define the maximum 
exponent for the LSB as $\texttt{emax}_{lsb} = \texttt{emax} - P + 1$. 

\textbf{Validity and canonical values}. We define values to be of type
\texttt {p3109}, which is an inductive type indexed by a format $f$.
\begin{lstlisting}[language=Lean, caption={Definition of the \texttt{p3109} inductive type}]
inductive p3109 (f : p3109_format) where
  | p3109_infinity (h : f.d = .extended) (sign : Bool) (hm : sign=true → f.s=.signed)
  | p3109_nan
  | p3109_finite (m e : ℤ)
    (hm : f.s = .unsigned → 0 ≤ m)
    (hcan : @canonical_p3109 f ⟨m, e⟩)
\end{lstlisting}
The \texttt{p3109} type is categorized into \texttt{p3109\_infinity},
\texttt{p3109\_nan}, and \texttt{p3109\_finite}. For infinite values
of type \texttt{p3109\_infinity}, $f$ must be \texttt{Extended}, and
its signedness must also coordinate with the representable values.

For $(m, e) \in f$ to be \texttt{p3109\_finite}, it must be \textbf{canonical}
and satisfy the signedness constraint: $m$ must be non-negative when $f$ is 
\texttt{Unsigned}. A value $(m, e)$ is \textbf{canonical} if 
$|m| < 2^P$, $\texttt{emin}_{lsb} \le e$, and it is either \textit{normal} 
or \textit{subnormal}. A value is normal if $2^{P-1} \le |m|$ and 
$e \le \texttt{emax}_{lsb}$, with the additional constraint that 
the significand field must not conflict with special-value encodings 
when $e = \texttt{emax}_{lsb}$. A value is subnormal if 
$e = \texttt{emin}_{lsb}$ and $|m| < 2^{P-1}$.

Lastly, we define a mapping $[\![\cdot]\!]$ from the P3109
algebraic model to the set of closed extended reals. We express closed
extended reals as a sum type of extended reals for finite values and
the unit type for special values. Finite P3109 values defined by the
pair $(m, e)$ map to $m \times 2^e$. For \texttt{Extended} formats,
$+\infty$ and $-\infty$ map to $\top$ and $\bot$
respectively. \texttt{NaN} maps to the unit, right injected.  We
formally define the value set $\mathcal{V}_{f}$ of a format $f$ as the
range of the evaluation function: $ \mathcal{V}_f = \{ v \mid \exists a
\in \texttt{p3109}_f, [\![a]\!] = v \} $.
We establish the invertibility of the mapping in the next section,
in which we prove the correctness of the formalization.

\subsection{Ensuring Model Fidelity: The Three-Way Isomorphism}
\label{sec:correctness}

We now establish the isomorphism sketched in 
Figure~\ref{fig:correctness-triangle}.

\begin{wrapfigure}{r}{.5\textwidth}
  \centering
  \includegraphics[width=\linewidth]{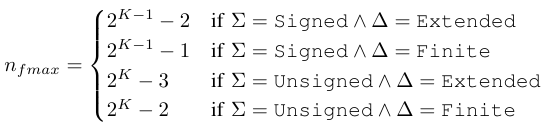}
  \caption{
  \small The definition of $n_{fmax}$, the encoding of the maximum 
  finite value in a given format.}
  \label{fig:max-finite}
\end{wrapfigure}

\textbf{Validating algebraic bounds}.
To establish the full isomorphism, we must first verify that the
algebraic bounds defining our ADT are mathematically sound. The most
delicate bound is the maximum exponent. As discussed in
Section~\ref{sec:formalization}, formats with precision $P \le 2$
reserve the highest exponent field encodings for special values,
dynamically lowering the maximum representable finite exponent. We
decode the exponent directly from the maximum finite number encoding
$n_{fmax}$ (Figure~\ref{fig:max-finite}) and prove its equality with
the \texttt{emax} defined in Figure~\ref{fig:emax} under case analysis
on the format parameters.

\begin{figure}[h]
  \centering
  \includegraphics[width=.8\linewidth]{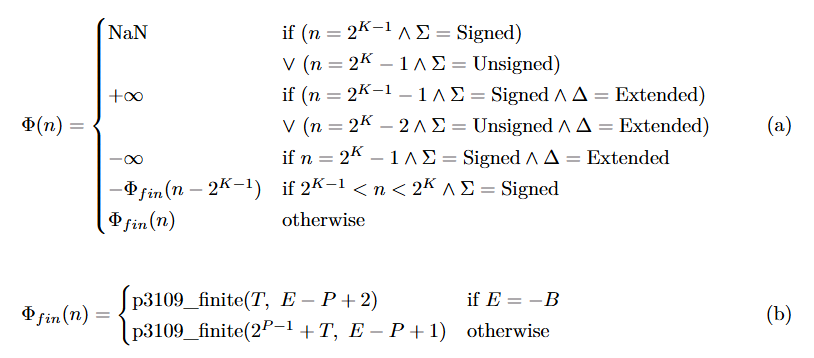}
  \caption{
  \small 
  (a) The bit-to-ADT decoding function.
  (b) The helper function for finite value encodings. 
  }
  \label{fig:phi}
\end{figure}

\textbf{Bijection between bit-level encodings and the ADT}.
We define the bit-to-ADT decoding function $\Phi : \text{Fin}(2^K)
\rightarrow \texttt{p3109}_f$ as a piecewise mapping
(Figure~\ref{fig:phi}a). For integers in the finite domain, $\Phi$
extracts the significand ($m$) and canonical exponent ($e$) via
Euclidean division. Given a target precision $P$ and exponent bias
$B$, let the trailing significand $T$ and the biased exponent $E$ be
defined as $T = n \bmod 2^{P-1}$ and $E = \lfloor n / 2^{P-1} \rfloor
- B$ respectively. The helper function $\Phi_{fin}$
(Figure~\ref{fig:phi}b) maps these components to the algebraic finite
constructor while distinguishing between zero, subnormal, and normal
values.

We prove that $\Phi$ is a bijection.

\begin{lemma}[$\Phi$ is injective]\label{lemma:phi-inj}
$\forall n_1, n_2 \in \text{Fin}(2^K), \quad \Phi(n_1) = \Phi(n_2) \implies n_1 = n_2$
\end{lemma}

\begin{lemma}[$\Phi$ is surjective]\label{lemma:phi-surj}
$\forall b : \texttt{p3109}_{f}, \quad \exists a \in \text{Fin}(2^K) \quad \textit{s.t.} \quad \Phi(a) = b$
\end{lemma}

A challenge with these proofs, which distinguishes it from the
corresponding IEEE-754 counterparts, is that P3109 dynamically shifts
the boundary between finite and special-value encodings based on the
format parameters $(K, \Sigma, \Delta)$, unlike IEEE-754's static
partition at $E = 2^W - 1$ (Table~\ref{tab:special_values}).  For
injectivity, we decompose the encoding space into disjoint regions and
use the uniqueness of Euclidean division within the finite region.
For surjectivity, we construct explicit bit-pattern witnesses for each
ADT constructor and verify they do not collide with adjacent
partitions.  Because P3109 shifts partition boundaries dynamically,
each format configuration requires separate verification---our
mechanization formally certifies that the ADT invariants correctly
partition the full encoding space.

\textbf{Differential testing}. The three-way isomorphism establishes
internal consistency of our model.  To additionally validate that it
faithfully captures the P3109 standard, we extracted the encode/decode
functions in the isomorphism and performed differential testing
against the value tables available in the standard.

\textbf{Semantic soundness}.  To connect the algebraic model to the
semantic domain, we define an encoding function $\texttt{encode} :
\mathcal{V}_f \to \texttt{p3109}_f$ that maps values back to the
algebraic structure and establish the round-trip identity:

\begin{theorem}[Round-Trip Identity]\label{theorem:round-trip}
$\forall a \in \texttt{p3109}_f, \ \texttt{encode}([\![a]\!]) = a$
\end{theorem}

Theorem~\ref{theorem:round-trip} confirms that $[\![\cdot]\!]$ is
injective and \texttt{encode} is its left inverse. Since
$[\![\cdot]\!]$ is surjective onto $\mathcal{V}_f$ by
construction, it is a bijection. Composing $\Phi$
and $[\![\cdot]\!]$ yields the full isomorphism between
$\text{Fin}(2^K)$ and $\mathcal{V}_f$.

\subsection{Projection}
\label{sec:rounding}

Floating-point operations in the P3109 standard are defined
mathematically on the set of closed extended reals and their results
are projected back into a representable value in the target
format. This projection is a three-step process: \textit{rounding} a
real number to a precision $P$ with unbounded exponents,
\textit{saturating} based on the format's finite range, and
\textit{encoding} the closed extended real back to a P3109 value.  The
standard defines a projection specification $\pi$ as a pair consisting
of a rounding mode and a saturation mode: $\pi = (Rnd_{\pi},
Sat_{\pi})$

\textbf{Rounding}. The \texttt{RoundToPrecision} function converts a
real number $x$ to a value representable with precision $P$ and an
unbounded exponent. P3109 supports all IEEE-754 rounding modes (\eg,
$\rne$, $\rz$) as well as $\ro$ and three variants of $\sr$. Our
implementation is parameterized by a rounding operator that rounds the
scaled significand of $x$ to an integer; for each operator, we provide
an axiomatic definition and prove the correctness of the
implementation. For the standard IEEE-754 modes, we formalize their
defining properties (\eg, $\rne$ returns the nearest neighbor and even
significand when at midpoint, $\rd$ returns the predecessor, $\ru$
returns the successor) and show they lift to the P3109 projection
pipeline. For the three $\sr$ variants, we formally verify their bias
properties~\cite{Fitzgibbon:stochastic:arith:2025}: the fastest
variant has exact bias $-2^{-(N+1)}$ for $N$ random bits, the second
variant is pointwise unbiased, and the last variant remains unbiased
even for finite-precision inputs.  We also prove that the unbiased
variants achieve $O(\sqrt{n})$ root-mean-square error over $n$
independent roundings.

We separate the unified typeclass that the operators must satisfy in 
Flocq~\cite{boldo:flocq:arith:2011}, a library that formalizes floating-point 
arithmetic in Rocq, into two typeclasses Faithful and 
Monotonic. While all the P3109 rounding modes are faithful, the stochastic 
rounding operators are not monotonic. Using only faithful rounding, however, we 
can still establish a weaker version of monotonicity where one side of the 
inequality is already representable. In fact, this weaker monotonicity property 
is sufficient to complete arithmetic proofs in the following sections.

To accommodate P3109's low-precision formats, our rounding operator has 
signature \texttt{rnd}: $\mathbb{Z}\to\mathbb{R}\to\mathbb{Z}$, where the 
first argument is the canonical exponent. This diverges from 
Flocq's~\cite{boldo:flocq:arith:2011} context-insensitive 
$\mathbb{R}\to\mathbb{Z}$ operator, which cannot implement P3109's 
exponent-dependent tie-breaking for $P=1$ formats (\eg, $\rne$ relies on 
\textit{exponent parity} rather than significand parity). 
We avoid partially applying the exponent, as doing so would make the operator 
binade-dependent and prevent the proof of global monotonicity.

\textbf{Saturation}. Using the saturation modes described in
Section~\ref{sec:background}, the \texttt{Saturation} function handles
values exceeding the format's finite range $[M^{lo}, M^{hi}]$. The
output of applying \texttt{RoundToPrecision} and \texttt{Saturation}
must be a value in the target value set. We prove this through case
analysis on whether the original input value is already in the finite
range. If it is, the result of \texttt{RoundToPrecision} and
\texttt{Saturation} is guaranteed to be in the value set due to the
aforementioned monotonicity property. For values greater than
$M^{hi}$, we prove that the saturated result is either $M^{hi}$ or
$+\infty$. Similarly, we prove that values smaller than $M^{lo}$
saturate to either $M^{lo}$ or $-\infty$.

\textbf{Encode}.  \texttt{Encode} takes a closed extended real value
$x \in \mathcal{V}_{f}$ and returns the corresponding \texttt{p3109}
algebraic value; its correctness is guaranteed by the bijection
established in the three-way isomorphism
(Theorem~\ref{theorem:round-trip}).

\textbf{Projection}. We assemble the three steps to form the
\texttt{Projection} pipeline.  We present key properties of
\texttt{Projection} that are relevant to our analysis in
Sections~\ref{sec:fasttwosum} and~\ref{sec:scalar}.

\begin{lemma}[Projection equals to rounding if already in range]\label{lemma:project-identity}
If an extended real value $r$ is in the finite range $M^{lo}\le r\le M^{hi}$
then \texttt{Projection} and \texttt{RoundToPrecision} produce the same outputs 
in the extended real domain.
\end{lemma}

\begin{lemma}[Projection is ``faithful'']\label{lemma:project-faith}
Projecting a real value $r$ under any $(Rnd_{\pi}, Sat_{\pi})$ produces $r$ if $r\in\mathcal{V}_f$, 
$pred(r)$ or $succ(r)$ otherwise.
\end{lemma}

Both lemmas follow by case analysis on whether $r$ is in the finite 
range $[M^{lo}, M^{hi}]$; out-of-range values saturate to boundary values 
that are themselves faithful roundings.

\textbf{Interaction of round-to-odd with saturation}.
A key property of $\ro$ is the ``even implies exact'' diagnostic: an
even significand in the result certifies that no rounding occurred,
which is vital for double-rounding
avoidance~\cite{park:odd-f2s:arxiv:2026}.  We formalized the
conditions under which this holds end-to-end through P3109's
round-then-saturate pipeline. For values within the representable
finite range $M^{lo} \le x \le M^{hi}$, saturation is not
triggered~(Lemma~\ref{lemma:project-identity}), and the diagnostic
holds universally.  However, when overflow triggers saturation, the
diagnostic's reliability depends on the parity of the boundary values
$M^{hi}$ and $M^{lo}$.  In configurations where these boundaries have
odd significands (e.g., \texttt{Signed Finite} and $P > 1$), the
diagnostic remains sound.  Conversely, for formats like
\texttt{Unsigned Finite} and \texttt{Signed Extended} ($P > 1$),
$M^{hi}$ has an even significand because the odd encoding is reserved
for special values. Saturated out-of-range inputs in these formats can
yield an even significand despite being inexact. Our mechanized proofs
provide a complete characterization: the diagnostic is sound for all
in-range values across all formats, for \texttt{Signed Finite} formats
under any saturation, and for \texttt{Unsigned Extended} formats under
any saturation mode, saturation can clip to a boundary value with an
even significand.

\section{Analysis of FastTwoSum for P3109}
\label{sec:fasttwosum}

\texttt{FastTwoSum} is a foundational error-free transformation (EFT) 
that computes the exact rounding error of floating-point addition. 
Introduced by Dekker~\cite{dekker:fts:numer-math:1971}, the algorithm 
computes a floating-point sum $s=\circ(a+b)$ and a correction term $t$. Under any 
round-to-nearest ($\rn$) mode, if $e_a \ge e_b$, it is guaranteed that 
$s+t = a+b$ exactly. This property makes it ubiquitous in algorithms 
that emulate twice the working precision or require intermediate error 
compensation~\cite{priest:arbitrary-prec:arith:1991, hida:qd:arith:2001}.

\begin{figure}[ht]\label{fig:fts}
  \centering
  \begin{subfigure}{0.35\linewidth}
    \begin{empheq}[box=\fbox]{align*}
    \textbf{Fa}&\textbf{stTwoSum}(a, b):\\
      &s = \circ(a+b)\\
      &z = \circ(x-a)\\
      &t = \circ(b-z)\\
      &\textbf{return } s, t
    \end{empheq}
    \caption{\centering Original \texttt{FastTwoSum} algorithm}
    \label{fig:fts-orig}
  \end{subfigure}
  \begin{subfigure}{0.45\linewidth}
    \includegraphics[width=\linewidth]{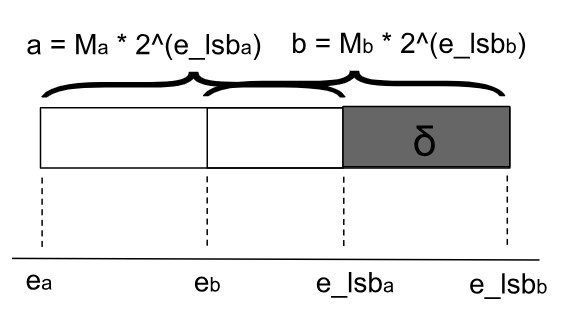}
    \caption{\centering Visualization of $\delta = a + b - s$}
    \label{fig:fts-error}
  \end{subfigure}
  \caption{
  (a) The expression $\circ(\cdot)$ denotes a floating-point operation that 
  rounds the real arithmetic result via a given rounding mode $\circ$ (\eg, 
  $\rne$). The two outputs $s$ and $t$ each represent the floating-point sum of 
  $a+b$ and an approximation of its associated rounding error 
  $\delta = a+b - s$.  
  (b) Suppose $a = M_{a} \times 2^{e\_lsb_{a}}$ and $b = M_{b} \times 2^{e\_lsb_
  {b}}$. Assuming $e\_lsb_{a} \le e_{b} \le e_{a}$ , 
  $s = \circ(a+b)$ rounds away the trailing bits of $b$. The bits in the gray 
  region form the rounding error $\delta = a+b - s$. If this region is exactly 
  representable with $P$-bits, $\delta$ is a floating-point number in the same 
  format as $a$ and $b$.
  }
\end{figure}

Given the widespread use of \texttt{FastTwoSum}, we developed a mechanized 
formalization for P3109 to account for its extremely low precision ($P=1$) 
and explicit saturation semantics, which prior IEEE-754 analyses exclude
~\cite{boldo:fast2sum-faithful:toms:2017, jeannerod:fts-revisited:arith:2025, 
boldo:flocq:arith:2011}. Using projections, we define \texttt{FastTwoSum} 
for finite P3109 inputs $a$ and $b$ as follows.

\begin{definition}[\texttt{FastTwoSum}]\label{def:fts}
Given two inputs $a, b \in \mathcal{V}_{f}\setminus\{\pm\infty, \texttt{NaN}\}$ such that $f$ is \texttt{Signed}, \texttt
{FastTwoSum} computes two outputs $s, t \in \mathcal{V}_{f}$ using three operations. Each 
operation is associated with projection specifications $(Rnd_1, Sat_1), (Rnd_2, 
Sat_2)$, and $(Rnd_3, Sat_3)$ respectively.
\begin{center}
\begin{minipage}{.5\linewidth}
\begin{align*}
    s &\leftarrow \Project_{(Rnd_1, Sat_1)} (a + b) \\
    z &\leftarrow \Project_{(Rnd_2, Sat_2)}(s - a) \\
    t &\leftarrow \Project_{(Rnd_3, Sat_3)}(b - z)
\end{align*}
\end{minipage}
\end{center}
\end{definition}

The primary objective of \texttt{FastTwoSum} is to ensure that $t$ is an 
accurate estimate of the error $\delta = a+b -s$. To analyze the accuracy of 
$t$ under P3109 arithmetic, we first establish the conditions under which the 
second operation is exact (\ie, no rounding error).

\begin{lemma}[Exact Intermediate Step]\label{lemma:exact-z}
Let $a, b \in \mathcal{V}_{f}\setminus\{\pm\infty, \texttt{NaN}\}$ for a \texttt{Signed} format $f$. If $|a+b|
\le M_{hi}$ and $e_a \ge e_b$, then $s - a \in \mathcal{V}_{f}$.
\end{lemma}

The proof for the above lemma can be found in~\cite[Lemma 2.5]{boldo:fast2sum-faithful:toms:2017}. 
Our formalization of this proof certifies that 
Lemma~\ref{lemma:exact-z} holds for all \texttt{Signed} P3109 formats, 
irrespective of the available precision $P$ and projections specifications
$(Rnd_1, Sat_1)$ and $(Rnd_2, Sat_2)$. Due to 
Lemma~\ref{lemma:project-faith}, which ensures projections preserve 
faithful rounding, Lemma~\ref{lemma:exact-z} induces 
$z = \Project_{(Rnd_3, Sat_3)}(s - a) = s - a$. Subsequently,  
$t =  \Project_{(Rnd_3, Sat_3)}(b - (s - a)) = \Project_{(Rnd_3, Sat_3)}
(a+b - s)$: $t$ is the projection of $\delta = a+b - s$ via 
$Rnd_3$ and $Sat_3$. As previously mentioned, if $s$ is a floating-point number 
\textit{nearest} to $a+b$, then $\delta$ is also a floating-point number. 
We establish this property for all \texttt{Signed} P3109 formats. 

\begin{lemma}[Representable Error]\label{lemma:delta-in-f}
Let $a, b \in \mathcal{V}_{f}\setminus\{\pm\infty, \texttt{NaN}\}$ for a \texttt{Signed} format $f$. If $|a
+b| \le  M_{hi}$ and $s = Project_{(Rnd_{1}, Sat_{1})}(a+b)$ for $Rnd_{1} \in \rn$, 
then $\delta = a+b - s \in \mathcal{V}_{f}$.
\end{lemma}

Note that Lemma~\ref{lemma:delta-in-f} only requires $Rnd_1$ to perform 
round-to-nearest (\ie, $Rnd_1 \in \rn$) without any specified 
tie-breaking rule. Because $Rnd_1$ is not restricted to $\rne$, we are able 
to prove Lemma~\ref{lemma:delta-in-f} for all \texttt{Signed} formats without 
separately addressing cases where $P = 1$. Since Lemma~\ref{lemma:exact-z} 
ensures $t$ is a projection of $\delta$ and projection preserves 
faithful rounding, Lemmas~\ref{lemma:exact-z} and~\ref{lemma:delta-in-f} 
collectively induce the following property.

\begin{theorem}[Exact Error Computation]\label{theorem:fts-eft}
Let $a, b \in \mathcal{V}_{f}\setminus\{\pm\infty, \texttt{NaN}\}$ for a \texttt{Signed} format $f$. If $|a+b| 
\le  M_{hi}$, $e_a \ge e_b$, and $Rnd_1 \in \rn$, then $t = a + b - s$.
\end{theorem}

Theorem~\ref{theorem:fts-eft} implies that under $\rne$, $|a+b| \le M_{hi}$ 
and  $e_a \ge e_b$ guarantee $s + t = a+b$ and that \texttt{FastTwoSum} is an 
EFT in P3109 arithmetic. We extend this analysis to the general case where 
the rounding mode is not $\rne$, a scenario relevant to P3109's 
support for diverse rounding modes. Previous analyses have established that 
even when $s$ is \textit{not} the nearest neighbor of $a+b$, $t$ is still 
a faithful rounding of $\delta$ so long as $z = s - a$~\cite{boldo:fast2sum-faithful:toms:2017}~\cite{jeannerod:fts-revisited:arith:2025}. 
As previously explained, Lemma~\ref{lemma:exact-z} induces $z = s - a$ and 
$t = \Project_{(Rnd_3, Sat_3)}(\delta)$. Through 
Lemma~\ref{lemma:project-faith}, which guarantees the faithfulness of 
projections, we establish that $t$ is a faithful rounding of 
$\delta$ for all \texttt{Signed} P3109 formats irrespective of the rounding 
mode used to compute $s$.

\begin{theorem}[Faithful Error Computation]\label{theorem:fts-faith}
Let $a, b \in \mathcal{V}_{f}\setminus\{\pm\infty, \texttt{NaN}\}$ for a \texttt{Signed} format $f$. If $|a+b| 
\le  M_{hi}$  and $e_a \ge e_b$, then $t$ is a faithful rounding of $\delta = a 
+ b - s$.
\end{theorem}

To derive $t = \Project_{(Rnd_3, Sat_3)}(b - z) = \Project_{(Rnd_3,
  Sat_3)} (a+b - s)$ from Lemma~\ref{lemma:exact-z},
Theorems~\ref{theorem:fts-eft} and~\ref{theorem:fts-faith} must ensure
that $z$ is a finite value (\ie, $|z| \le M_{hi}$). Note that
Theorems~\ref{theorem:fts-eft} and~\ref {theorem:fts-faith} both
require $|a+b| \le M_{hi}$, which subsequently ensures $|s| \le
M_{hi}$ (\ie, the first operation does not overflow to infinity).
Therefore, we must establish that $|s| \le M_{hi}$ guarantees $|s - a|
\le M_{hi}$ and the second operation $z \leftarrow \Project_{(Rnd_2,
  Sat_2)} (s - a)$ does not induce overflow. Boldo \etal prove this
property in~\cite [Theorem
  5.1]{boldo:fast2sum-faithful:toms:2017}. However, the proof requires
the target format to have at least 3 bits of precision. Through new
proofs that separately address $P = 1$ and $P = 2$, we confirm that
\texttt{FastTwoSum} is immune to any intermediate overflow for all
\texttt{Signed} formats regardless of their precision.

\begin{theorem}[Overflow Immunity]\label{theorem:overflow-immune}
Let $a, b \in \mathcal{V}_{f}\setminus\{\pm\infty, \texttt{NaN}\}$ for a \texttt{Signed} format $f$. 
If the operation $s \leftarrow \Project_{(Rnd_1, Sat_1)}(a + b)$ does not
induce overflow, then the operation $z \leftarrow \Project_{(Rnd_2, Sat_2)}
(s - a)$ does not induce overflow.
\end{theorem}

The properties $s - a \in \mathcal{V}_{f}$ (Lemma~\ref{lemma:exact-z}) and $\delta \in \mathcal{V}_{f}$
(Lemma~\ref{lemma:delta-in-f}) that underlie \texttt{FastTwoSum}'s EFT 
guarantees both require $s$ to be finite.  Theorem~\ref{theorem:fts-eft} thus 
assumes $|a+b| \le M_{hi}$ to ensure the first operation does not overflow to 
infinity regardless of $Rnd_1$ and $Sat_1$. Due to its explicit saturation 
semantics, addition in P3109 can ensure $|s| \le M_{hi}$ even when $|a+b| > M_{hi}$ (\eg, $Sat_{1} \in \{\texttt{SatFinite}, \texttt{SatPropagate} \}$ or $f$ 
is \texttt{Finite}). Based on this observation, we identify a novel property of 
\texttt{FastTwoSum} under P3109 arithmetic: if $|a+b| > M_{hi}$ and $|s| = M_{hi}$, $e_a \ge e_b$ guarantees $t = a+b - s$.

\begin{theorem}[Exact Overflow Error Computation]\label{theorem:overflow-exact}
Let $a, b \in \mathcal{V}_{f}\setminus\{\pm\infty, \texttt{NaN}\}$ for a \texttt{Signed} format $f$. If $|a+b| > M_{hi}$, 
$|s| = |M_{hi}|$, and $e_a \ge e_b$, then $t = a + b - s$.
\end{theorem}

Theorem~\ref{theorem:overflow-exact} implies that when the real result 
$a+b$ is in the overflow range and $s$ is finite, \texttt{FastTwoSum} can 
compute the \textit{exact overflow error}. Furthermore, 
Theorem~\ref{theorem:overflow-exact} holds irrespective of the rounding mode 
used to compute $s$. By addressing sums in the overflow range, 
Theorem~\ref{theorem:overflow-exact} broadens \texttt{FastTwoSum}'s applicable 
domain. Our findings thus highlight the utility of P3109's saturation 
features, as they preserve \texttt{FastTwoSum}'s EFT guarantees in settings 
excluded by pure rounding.

\section{Analysis of ExtractScalar}
\label{sec:scalar}

\texttt{FastTwoSum} is the foundation of various floating-point splitting 
algorithms, which decompose an input $x$ across two numbers $x_{h}$ and $x_
{\ell}$ such that $x = x_{h} + x_{\ell}$. The most prominent example is 
Dekker's splitting~\cite{dekker:fts:numer-math:1971}, which splits a $P$-precision number 
across two values each with at most $\lfloor \frac{P}{2} \rfloor$ effective 
bits of precision (\ie, available precision minus the number of trailing 
zeroes). When applied iteratively, splitting algorithms enable floating-point 
expansions, which represent numbers across multiple non-overlapping, 
lower-precision values. Given the low-precision nature of P3109 formats, 
formalizing floating-point splitting techniques for P3109 representations 
is of practical use to the design of future algorithms.

\begin{figure}[t]\label{fig:extract-scalar}
  \centering
  \begin{subfigure}{0.45\linewidth}
    \includegraphics[width=\linewidth]{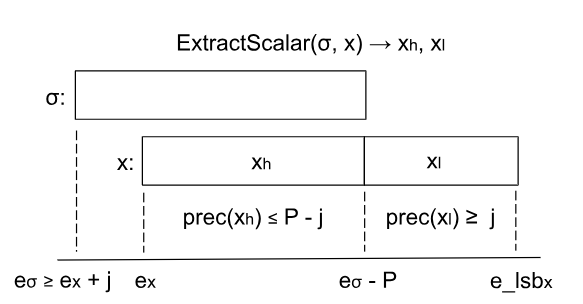}
    \caption{\centering Visualization of ExtractScalar}
    \label{fig:extract-scalar-ex}
  \end{subfigure}
  \begin{subfigure}{0.5\linewidth}
    \includegraphics[width=\linewidth]{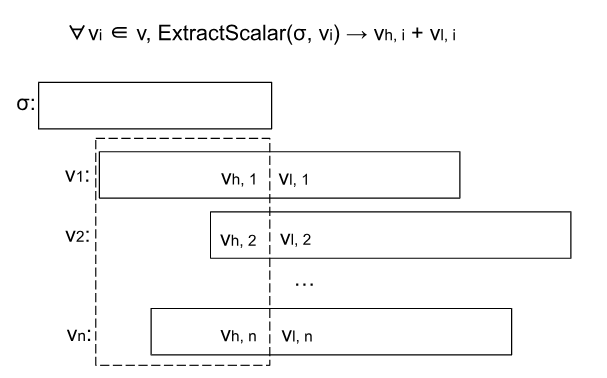}
    \caption{\centering Application of ExtractScalar}
    \label{fig:extract-vector-ex}
  \end{subfigure}
  \caption{
  (a) \texttt{ExtractScalar}'s outputs $x_{h}$ and $x_{\ell}$ satisfy $x = x_
  {h} + x_{\ell}$. Given $j \in \mathbb{Z}$ such that $|x| \le 2^{-j}\sigma$, 
  the floating-point sum of $\sigma$ and $x$ rounds away at least $j$ trailing
  bits of $x$, which are captured in $x_{\ell}$. $x_{h}$, which captures 
  $x$'s leading bits, has up to $P - j$ effective precision bits.
  (b) Given a floating-point vector $v \in f^{n}$, applying 
  \texttt{ExtractScalar} on each element $v_{i}$ produces vectors 
  $v_{h}, v_{\ell} \in f^{n}$ such that $v = v_{h} + v_{\ell}$. 
  The elements of $v_{h}$ collectively operate as \textit{fixed-point} numbers.
  }
\end{figure}

Unlike Dekker's splitting, which assigns a fixed number of bits, 
\texttt{ExtractScalar}~\cite{rump:accsum-1:siam:2008} is an algorithm for which 
the distribution of precision is configurable. Using the projection 
specification, we define \texttt{ExtractScalar} for finite P3109 inputs as 
follows.

\begin{definition}[ExtractScalar]\label{def:extract-scalar}
Given two finite inputs $\sigma, x \in \mathcal{V}_{f}$ such that $f$ is \texttt{Signed},
\texttt{ExtractScalar} computes two outputs $x_{h}, x_{\ell} \in \mathcal{V}_{f}$ using 
three operations performed under $\rne$ and saturation modes $Sat_1$, $Sat_2$, and $Sat_3$.
\begin{center}
\begin{minipage}{.5\linewidth}
\begin{align*}
    &s \leftarrow \Project_{(\rne, Sat_1)}(\sigma + x) \\
    &x_{h} \leftarrow \Project_{(\rne, Sat_2)}(s - \sigma) \\
    &x_{\ell} \leftarrow \Project_{(\rne, Sat_3)}(x - x_{h})
\end{align*}
\end{minipage}
\end{center}
\end{definition}

\texttt{ExtractScalar} assumes $\sigma = 2^{i}$ for some $i \in \mathbb{Z}$
(\ie, $\sigma$ is an integer power of 2). \texttt{ExtractScalar} also assumes 
$|x| \le 2^{-j}\sigma$ for $j \in \mathbb{Z}$~($\ge 0)$ such that $\sigma + x \le M_{hi}$ 
(\ie, no overflow). Given these assumptions, Rump \etal prove that 
\texttt{ExtractScalar}'s outputs $x_{hi}$ and $x_{\ell}$ satisfy $x = x_{h} + x_
{\ell}$: \texttt{ExtractScalar} is an EFT of $x$~\cite{rump:accsum-1:siam:2008}. Moreover, they establish that the number of 
effective bits in $x_{h}$ and $x_{l}$ are determined by the working precision 
$P$ and the parameter $j$ used to define $\sigma$ (see Figure~\ref{fig:extract-scalar-ex}).

While producing an EFT, \texttt{ExtractScalar} 
enforces several bounds on $x_{h}$ and $x_{\ell}$: $|x_{h}| \le 2^{-j}\sigma$, 
$x_{h} \in 2^{-P}\sigma\mathbb{Z}$ (\ie, $x_{h}$ is an integer 
multiple of $2^{-P}\sigma$ for precision $P$), and $|x_{\ell}| \le 2^{-P}
\sigma$. Through $|x_{h}| \le 2^{-j}\sigma$ and $x_{h} \in 2^{-P}\sigma\mathbb
{Z}$, \texttt{ExtractScalar} ensures that the number of effective bits in 
$x_{h}$ is determined by the \textit{exponent difference} between $\sigma$ and 
$x$.
%
%
\texttt{ExtractScalar} thus allows $\sigma$ to be 
configured with the appropriate $j$ to achieve the desired distribution across 
$x_{h}$ and $x_{\ell}$. In conjunction with the property 
$|x_{l}| \le 2^{-P}\sigma$, $|x_{h}| \in 2^{-P}\sigma\mathbb{Z}$ also ensures 
that $x_{h}$ and $x_{\ell}$ are non-overlapping.

\texttt{ExtractScalar}'s guarantees of $|x_{h}| \le 2^{-j}\sigma$ and $x_{h} 
\in 2^{-P}\sigma\mathbb{Z}$ also enable EFTs for floating-point vectors
operations. Given a $n$-length floating-point vector $v \in \mathcal{V}_{f}^{n}$, suppose 
$\sigma$ is configured with respect to $\text{max}_{i}|v_{i}|$ (\ie, 
$\text{max}_{i}|v_{i}| \le 2^{-j}\sigma$). Applying \texttt{ExtractScalar} on 
each element $v_{i}$ with the given $\sigma$ subsequently produces vectors
$v_{h}, v_{\ell} \in \mathcal{V}_{f}^{n}$ such that $v = v_{h}+v_{\ell}$. Furthermore, 
\texttt{ExtractScalar} ensures that for all elements $v_{h, i}$, 
$|v_{h,i}| \le 2^{-j}\sigma$ and $v_{h, i} \in 2^{-P}\sigma\mathbb{Z}$.
With their effective bits occupying the same range 
(see Figure~\ref{fig:extract-vector-ex}), the elements of $v_{h}$ 
can be interpreted as fixed-point numbers. In~\cite{rump:accsum-1:siam:2008},
Rump \etal leverage this property to design an algorithm that adds all elements 
of $v_{h}$ without inducing any rounding errors.
This strategy has subsequently enabled accurate, vector-based algorithms 
for floating-point accumulation~\cite{rump:accsum-1:siam:2008}
\cite{rump:accsum-2:siam:2009} and dot products~\cite{ozaki:ozaki-scheme:numer-algo:2012}.

While Dekker's splitting is a well-studied algorithm that has been analyzed 
using mechanized proofs~\cite{boldo:veltkamp-proof:ijcar:2006}, 
\texttt{ExtractScalar} currently lacks such rigorous formalization. We have 
thus developed to our knowledge the first mechanized proof of 
\texttt{ExtractScalar} for P3109 formats. We present our key findings 
below.

\begin{theorem}[Extract Scalar Properties]
Let $\sigma, x \in \mathcal{V}_{f}\setminus\{\pm\infty, \texttt{NaN}\}$ for a 
\texttt{Signed} format $f$ such that $P > 1$. Let 
$\sigma = 2^{i}$ for some $i \in \mathbb{Z}$. Let $|x| \le 2^{-j}\sigma$ for 
some $j \in \mathbb{Z}$ such that $0 \le j$. If $|\sigma + x| \le M_{hi}$, 
$x_{h}$ and $x_{\ell}$ satisfy the following conditions.

\begin{center}
\begin{minipage}{.5\linewidth}
\begin{equation*}\label{eq:es-1}
    (a) \ x = x_{h} + x_{\ell}
\end{equation*}
\end{minipage}
\begin{minipage}{.5\linewidth}
\begin{equation*}\label{eq:es-2}
    (b) \ |x_{\ell}| \le 2^{-P}\sigma
\end{equation*}
\end{minipage}
\begin{minipage}{.5\linewidth}
\begin{equation*}\label{eq:es-3}
    (c) \ x_{h} \in 2^{-P}\sigma\mathbb{Z}
\end{equation*}
\end{minipage}
\begin{minipage}{.5\linewidth}
\begin{equation*}\label{eq:es-4}
    (d) \ |x_{h}| \le 2^{-j}\sigma
\end{equation*}
\end{minipage}
\end{center}

\end{theorem}

Note that $s$ and $x_{\ell}$ are the outputs of performing \texttt{FastTwoSum} 
(see Definition~\ref{def:fts}) on $\sigma$ and $x$ under $\rne$. 
Furthermore, \texttt{ExtractScalar}'s precondition $|x| \le 2^{-j}\sigma$ 
implies $e_{\sigma} \ge e_{x}$. As such, Lemma~\ref{lemma:exact-z} and 
Theorem~\ref{theorem:fts-eft} imply $x_{h} = s - \sigma$ and $x_{\ell} 
= \sigma + x - s$ respectively. Subsequently, $x = x_{h} + x_{\ell}$, and thus 
\texttt{ExtractScalar} is also an EFT under P3109 arithmetic. The core proofs 
for post conditions (b) through (d) are available in~\cite[Lemma 3.3, p.201]{rump:accsum-1:siam:2008}. 
A direct application of these proofs to our model confirms these conditions for 
all \texttt{Signed} P3109 formats \textit{with at least 2-bits of available 
precision}. 

The original proof of the last condition $|x_{h}| \le 2^{-j}\sigma$ leverages 
the properties of $\rne$. The proof asserts that when 
$\sigma + x$ is at the exact midpoint between $\sigma$ and one of its 
floating-point neighbors, $\sigma$ being an integer power of 2 induces 
$s = \sigma$ under $\rne$ (see first operation in Definition~\ref{def:extract-scalar}). The equality $s = \sigma$ induces $x_{h} = 0$, which 
would satisfy $|x_{h}| \le 2^{-j}\sigma$ for the relevant values of $j$. 
However, the proof implicitly assumes all integer powers of 2 in the target 
format have even significands, which is not the case for P3109 formats where 
$P=1$. As detailed in Section~\ref{sec:rounding}, the tie-breaking for 
$\rne$ when $P=1$ utilizes the parity of the exponent. As such, the original 
proof's requirement that the floating-point sum of $\sigma$ and $x$ rounds 
to $\sigma$ won't hold when $\sigma = 2^{i}$ for an odd integer $i$.
Consequently, our analysis indicates that condition (d) does not hold for 
$P = 1$.

\section{Lean Development}
We developed \tool using Lean 4. The entire development is
approximately fifteen thousand lines of code, covering the P3109
bit-encodings, the algebraic model, the isomorphism proofs, and the
analysis of \texttt{FastTwoSum} and \texttt{ExtractScalar}. The
development relies on \texttt{Mathlib}'s theory of extended real
numbers and serves as a mathematical specification rather than an
executable implementation.

\textbf{Errors discovered}. During the process of formalizing the
P3109 standard, we discovered multiple errors in the interim P3109
report. We reported them to the working group and they have been
fixed. We highlight few of the errors that we identified.
\begin{enumerate}
\item We discovered that the \texttt{emax} for unsigned formats was
  incorrectly specified. The interim P3109 report had indicated that
  $2^W -1 - B$ as the \texttt{emax} whereas it does not hold for
  representations with precision $P=1$ and $P=2$ as described in
  Section~\ref{sec:formalization}.

\item We identified that the \texttt{encode} function had an error in
  the specification of infinity. The interim report had erroneously
  indicated that $2^{K-1}-2$ as the infinity for the
  \texttt{unsigned}, \texttt{extended} format. We reported that it
  should be changed to $2^{K}-2$.

\item We also identified that a note in the report about projection
  was incorrect for $\ru$, which has been subsequently been corrected.

\end{enumerate}  

\section{Related Work}
\label{sec:related}

The formalization of floating-point
arithmetic~\cite{muller:fp-handbook:2009} has a rich history spanning
multiple proof assistants.
One of the earliest efforts was 
Harrison's~\cite{harrison:hol-light-fp:tphols:1999} formalization in HOL 
Light, which verified floating-point algorithms such as the exponential 
function. In PVS, Boldo and 
Mu\~noz~\cite{boldo:pvs-fp:nasa:2006} developed a parametric formalization 
that abstracts over format parameters. In Coq, the PFF 
library~\cite{daumas:pff:TPHOLs:2001, boldo:veltkamp-proof:ijcar:2006, 
boldo:underflow:arith:2003} provided a high-level, axiomatic formalization 
of IEEE-754, but relied on rounding defined as a relation rather than a 
computable function, limiting its use for executable specifications. 
Flocq~\cite{boldo:flocq:arith:2011} addressed this by unifying PFF's 
axiomatic strengths with executability via the ``Generic Format'', and 
remains the state-of-the-art for IEEE-754 verification. Flocq is actively 
being ported to the 
\texttt{FloatSpec}~\cite{beneficial-ai:floatspec:2026} library in Lean. 
In Isabelle, Yu~\etal~\cite{yu:ieee754-isabelle:isa-afp:2013} formalize 
IEEE-754 with a focus on executable representations.
However, none of these frameworks can be directly adapted for P3109 due to 
structural mismatches in their rounding operator signatures and lack of 
saturation semantics (see Section~\ref{sec:rounding}).

Closest to our work,
Wintersteiger~\etal~\cite{Wintersteiger:imandra-p3109:arith:2025} formalized 
the operational specification of P3109 using Imandra, producing an executable 
reference implementation with automatically extracted OCaml code. Their 
verification focuses on totality, proving that each operation always produces 
a valid P3109 value rather than an error. Our work differs in scope: we 
establish a three-way isomorphism between bit-encodings, an algebraic model, 
and the semantic domain; we formalize the full projection pipeline including 
all rounding and saturation modes; and we analyze the correctness properties 
of numerical algorithms (\texttt{FastTwoSum} and \texttt{ExtractScalar}) 
within P3109 arithmetic.
SMT solvers such as Z3~\cite{demoura:z3:tacas:2008} support IEEE-754 
floating-point theory but do not model P3109's parametric formats or 
saturation modes.

Rounding-error analysis tools such as 
Gappa~\cite{deDinechin:gappa:trans-comput:2011} and 
VCFloat2~\cite{appel:vcfloat2:cpp:2024} generate verification conditions for non-overflow but are designed around 
IEEE-754's overflow-to-infinity semantics. 
P3109's \textit{saturation arithmetic} treats overflow as a valid state 
that clamps to finite bounds, a behavior outside the scope of these tools.
The RLibm project~\cite{lim:rlibm-all:popl:2022, park:rlibm:pldi:2025} 
has recently analyzed \texttt{FastTwoSum} under 
round-to-odd~\cite{park:odd-f2s:arxiv:2026}, which is directly relevant 
to P3109 as it adopts round-to-odd as a rounding mode.
Other efforts in correctly rounded libraries~\cite{Daramy:crlibm:spie-5205:2003, sibidanov:core-math:arith:2022}, 
automated error optimization~\cite{panchekha:herbie:pldi:2015, darulova:daisy:tacas:2018}, 
and safety-critical analysis~\cite{titolo:precisa:fm:2024} address 
complementary aspects of floating-point verification but do not target 
P3109's parametric formats or saturation semantics.

\section{Conclusion}
\label{sec:conclusion}

The \tool framework provides a machine-checked specification of the
upcoming IEEE P3109 standard for low-precision representations in Lean
while addressing the challenges created by new features such as
stochastic rounding and saturation. We show a three-way equivalence
between the bit-level encodings, our algebraic inductive type, and the
semantic value set of closed extended reals. We show the usefulness of
\tool by analyzing foundational FP algorithms such as
\texttt{FastTwoSum} and the FP splitting technique
\texttt{ExtractScalar}.  Our verification of \texttt{FastTwoSum}
revealed a novel property specific to P3109: under saturation, the
algorithm computes the exact ``overflow error'' regardless of the
rounding mode.  Our analysis for \texttt{ExtractScalar} shows that
previously known properties fail for formats with one bit of
precision, highlighting the necessity of bespoke verification for
ultra-low precision domains.

As P3109-compliant hardware accelerators emerge, \tool's bit-level 
isomorphism provides a natural bridge for RTL equivalence checking: 
properties proven on our algebraic model transfer directly to the 
concrete bit-encodings that hardware implements. The verified projection 
pipeline---including saturation and all rounding modes---can serve as a 
golden reference model against which hardware implementations are 
validated. Future work includes extending the algorithm analyses to 
\texttt{Unsigned} formats, which are relevant for machine learning 
operations such as ReLU that produce only non-negative values, and 
connecting \tool to automated verification tools such as
SMT-LIB's floating-point theory.

\section*{Acknowledgments}
  We thank the P3109 working group members especially Jeffrey Sarnoff
  and Christoph Wintersteiger for encouraging us to explore a Lean
  formalization. We would like to thank Ilya Sergey for his
  encouragement to explore Lean during Santosh's visit to NUS. This
  material is supported in part by the research gifts from the Intel
  corporation and National Science Foundation grants: NSF-2110861 and
  NSF-2312220. 
  Any opinions, findings, and conclusions or recommendations expressed
  in this material are those of the authors and do not necessarily
  reflect the views of the Intel corporation or the National Science
  Foundation.

\bibliography{reference}

\end{document}